\documentclass[11pt,a4paper]{scrartcl}

\usepackage{ILD}

\usepackage[symbol]{footmisc}
\usepackage{feynmf}
\usepackage{multirow}
\usepackage{textpos}

\title{Tuning Pythia8 for future $e^+e^-$ colliders}

\ildproc{PHYS}{2023}{002} 

\date{DESY-23-111}

\addauthor{Zhijie Zhao}{\institute{1,2}\thanks{Talk presented at the International Workshop on Future Linear Colliders (LCWS 2023), 15-19 May 2023. C23-05-15.3.}}
\addauthor{Mikael Berggren}{\institute{1}}
\addauthor{Jenny List}{\institute{1}}

\addinstitute{1}{Deutsches Elektronen-Synchrotron DESY, Germany}
\addinstitute{2}{Center for Future High Energy Physics, Institute of High Energy Physics, Chinese Academy of Sciences, China}

\abstract{
The majority of Monte-Carlo (MC) simulation campaigns for future $e^+e^-$ colliders has so far been based on the leading-order (LO) matrix elements provided by Whizard 1.95, 
followed by parton shower and hadronization in Pythia6, 
using the tune of the OPAL experiment at LEP. 
In this contribution, we test and develop the interface between Whizard3 and Pythia8. 
As a first step, we simulate the $e^+e^-\to q\bar{q}$ process with LO matrix elements, 
and compare three tunes in Pythia8: the standard Pythia8 tune, the OPAL tune and the ALEPH tune. 
At stable-hadron level, predictions of charged and neutral hadron multiplicities of these tunes are compared to LEP data, 
since they are strongly relevant to the performance of particle flow algorithms.

The events are used to perform a full detector simulation and reconstruction of the International Large Detector concept (ILD) as an example for a particle-flow-optimised detector. 
At reconstruction level, a comparison of the jet energy resolution in these tunes is presented. 
We found good agreement with previous results that were simulated by Whizard1+Pythia6. 
In addition, the preliminary next-to-leading order (NLO) results are also presented. 
This modern MC simulation chain, with matched NLO matrix elements in the future, should be introduced to ILC or other future $e^+e^-$ colliders.
}

\graphicspath{ {./logos/}{./figures/} }

\begin{document}

\titlepage






\section{Introduction}
\label{sec:intro}

To measure the properties of Higgs boson precisely, 
there are many proposals for future colliders, 
such as ILC~\cite{ILCInternationalDevelopmentTeam:2022izu,Behnke:2013xla,ILC:2013jhg,Adolphsen:2013jya,Adolphsen:2013kya,Behnke:2013lya}, 
CLIC~\cite{Linssen:2012hp,CLICdp:2018cto}, 
CEPC~\cite{CEPCPhysicsStudyGroup:2022uwl,CEPCStudyGroup:2018rmc,CEPCStudyGroup:2018ghi} 
and FCC-ee~\cite{Bernardi:2022hny,FCC:2018evy}. 
All of them are $e^+e^-$ colliders, 
and they are designed as Higgs Factories. 
There are many features of lepton beam: 
initial state radiation (ISR), polarization, beam-strahlung, and so on.
They also give special requirements to the Monte-Carlo (MC) event generator.
Whizard~\cite{Kilian:2007gr} is a general purpose event generator that can handle these features, 
and has been widely used in analyses of processes at $e^+e^-$ colliders.

There are many detector concepts for future Higgs Factories. 
In this contribution, we take International Large Detector (ILD)~\cite{Behnke:2013lya,ILDConceptGroup:2020sfq} as an example. 
ILD is designed for $e^+e^-$ collisions between 90 GeV and 1 TeV. 
It is optimized for particle flow algorithm (PFA). 
The PFA aims at recontructing every invidual particle created in the event, i.e., charged particles, photons, and neutral hadrons. 
In particular, the energy resolution of neutral hadrons is expected to be worse than that of the other two types of particle.
In the MC simulation, the reconstruction of these particles is strongly dependent on the tuning of MC parameters.

Some physics aspects cannot be derived from first principles, 
especially in the area of soft QCD. 
Pythia~\cite{Bierlich:2022pfr} is a MC generator that contains many parameters that represent a true uncertainty in our understanding of nature. 
Tuning these parameters is important to describe data. 
A good tune should have: 
\begin{itemize}
  \item Physically sensible parameter values, with good universality.
  \item Good agreement with data (LEP1 in this contribution).
  \item Reliable uncertainties.
  \item Best fit for our observables.
\end{itemize}

At present, events for analysis of $e^+e^-$ colliders have so far been based on the leading order (LO) matrix elements provided by WHIZARD 1.95. 
Parton shower and hadronization are performed by Pythia6~\cite{Sjostrand:2006za}, 
using the tune of the OPAL experiment at the Large Electron-Positron Collider (LEP). 
However, the major versions of Whizard and Pythia have been upgraded to Whizard3 and Pythia8. 
Our first goal is upgrading the simulation chain to Whizard3+Pythia8.
We hope to get good agreement with LEP data, especially the neutral hadrons. 
Finally, next-to-leading order (NLO) matched events should be included in the future, 
because of the requirement of high precision.

This contribution is organized as given below. 
In section~\ref{sec:ahm}, we compare the average hadron multiplicities of different Pythia8 and Pythia6 tunes. 
In section~\ref{sec:jer}, the full ILD simulation is performed and the jet energy resolution (JER) is discussed. 
Section~\ref{sec:nlo} shows the preliminary results of NLO events. 
Finally, we summarize this study in section~\ref{sec:summary}.

\section{Average Hadron Multiplicities}
\label{sec:ahm}

As we mention above, hadronization rates are crucial for studying particle flow performance. 
To study it, we use the following generator setup (LEP1 condition):
\begin{itemize}
     \item Process: $e^+e^-\to q\bar{q}$ $(q=u,d,s,c,b)$.  
     \item The center of mass energy is $E_{cm}=91.19$ GeV.
     \item Beams are un-polarized.
     \item Beam-strahlung is not considered.
     \item ISR is switched on. 
\end{itemize}
The parton shower and hadronization are performed by Pythia8. 
Three tunes are considered in this context: 
\begin{itemize}
  \item The standard tune, using the default parameter set of Pythia8.
  \item The OPAL tune. 
  \item The ALEPH tune.
\end{itemize}

\begin{figure}[htbp]
  \centering
  \begin{subfigure}[b]{0.45\linewidth}
     \includegraphics[width=\linewidth]{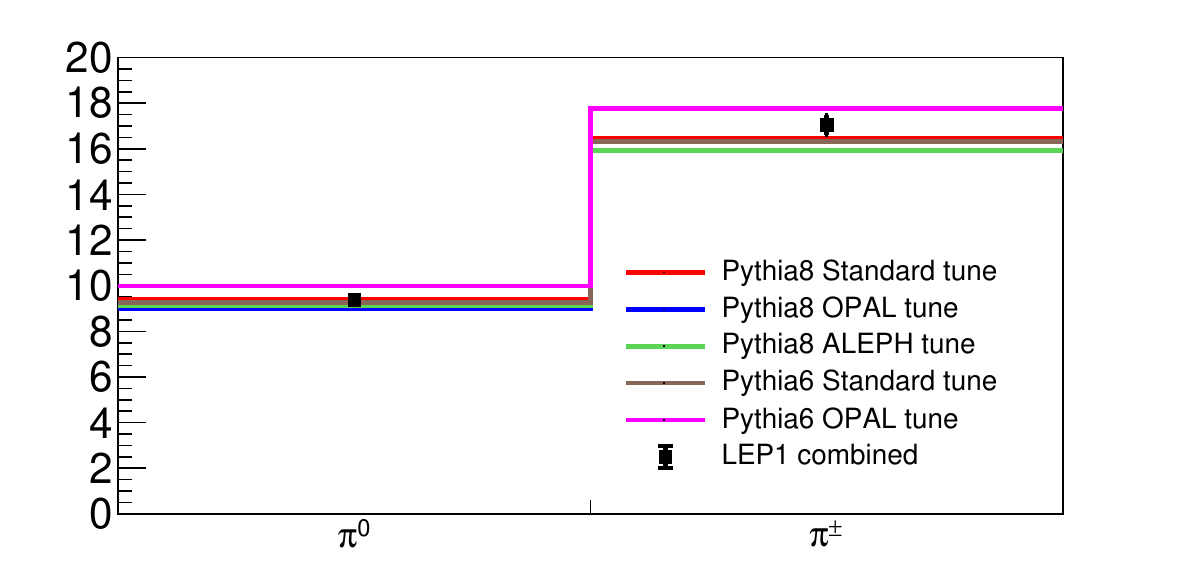}
  \end{subfigure}
  \begin{subfigure}[b]{0.45\linewidth}
     \includegraphics[width=\linewidth]{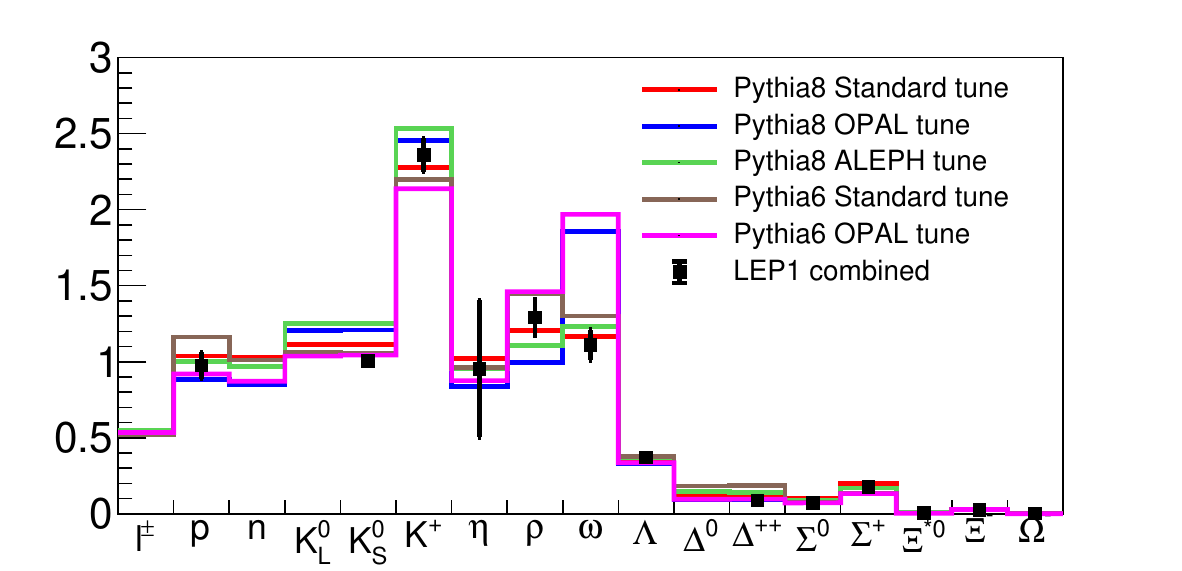}
  \end{subfigure}
  \caption{The average numbers of hadrons in $e^+e^-\to q\bar{q}$ events are displayed, with three Pythia8 tunes: the standard, OPAL and ALEPH tunes, and two Pythia6 tunes: the standard and OPAL tunes.} \label{fig:hadrons}
\end{figure}

The dominant hadrons are pions. 
The average numbers of $\pi^0$ and $\pi^\pm$ are displayed in the left of Fig.~\ref{fig:hadrons}.
Two Pythia6 tunes, standard and OPAL, are considered for comparison.
The LEP1 data are taken from \cite{Boehrer:1996pr,ALEPH:1996oqp} as a reference. 
We can see that three Pythia8 tunes and the Pythia6 standard tune are close to each other and generally agree with the LEP1 data, 
but the Pythia6 OPAL tune gives larger numbers.

The numbers of other hadrons are displayed in the right of Fig.~\ref{fig:hadrons}. 
For the proton and $K^0_S$, which are important in the calculation of JER, 
the usual Pythia6 OPAL tune has good agreement with LEP1 data, 
but the numbers of other hadrons have large differences to data. 
Generally, the standard Pythia8 tune is the closest one to data.

\section{ILD simulation and jet energy resolution (JER)}
\label{sec:jer}

Full Geant4-based MC simulations are crucial to optimize a well performing detector concept. 
We take ILD as an example to discuss the detector performance. 
In this context, an important parameter is the JER of ILD. 
To study it, we use the following generator setup: 
\begin{itemize}
  \item Process: $e^+e^-\to q\bar{q}$ $(q=u,d,s)$.
  \item ISR is switched off. 
  \item $E_{cm}=40,91,200,350,500$ GeV.
  \item Full simulation is performed with large model of ILD (ILD-L).
\end{itemize}

\begin{figure}[htbp]
  \centering
  \begin{subfigure}[b]{0.32\linewidth}
    \includegraphics[width=\linewidth]{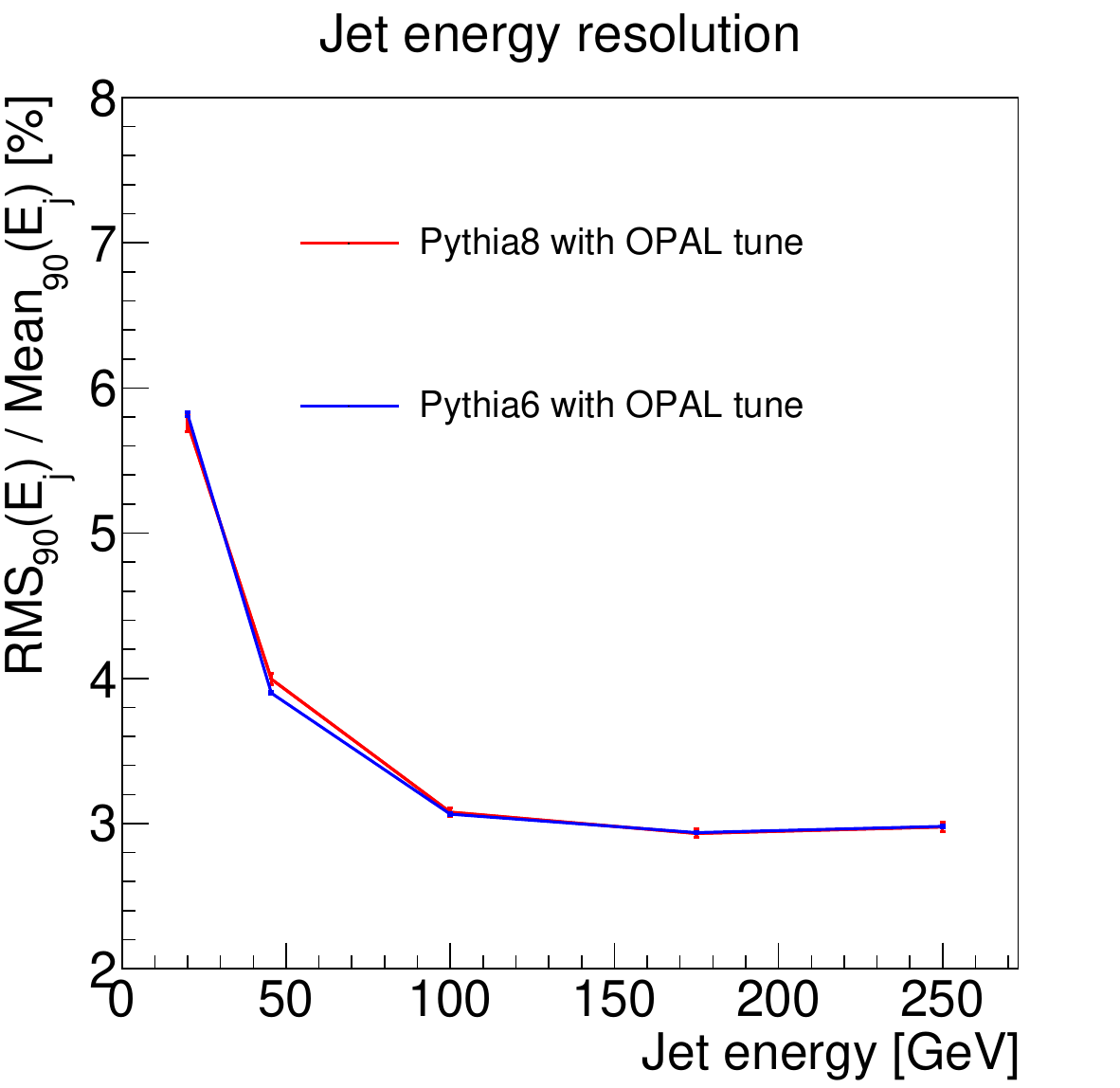}
  \end{subfigure}
  \begin{subfigure}[b]{0.32\linewidth}
    \includegraphics[width=\linewidth]{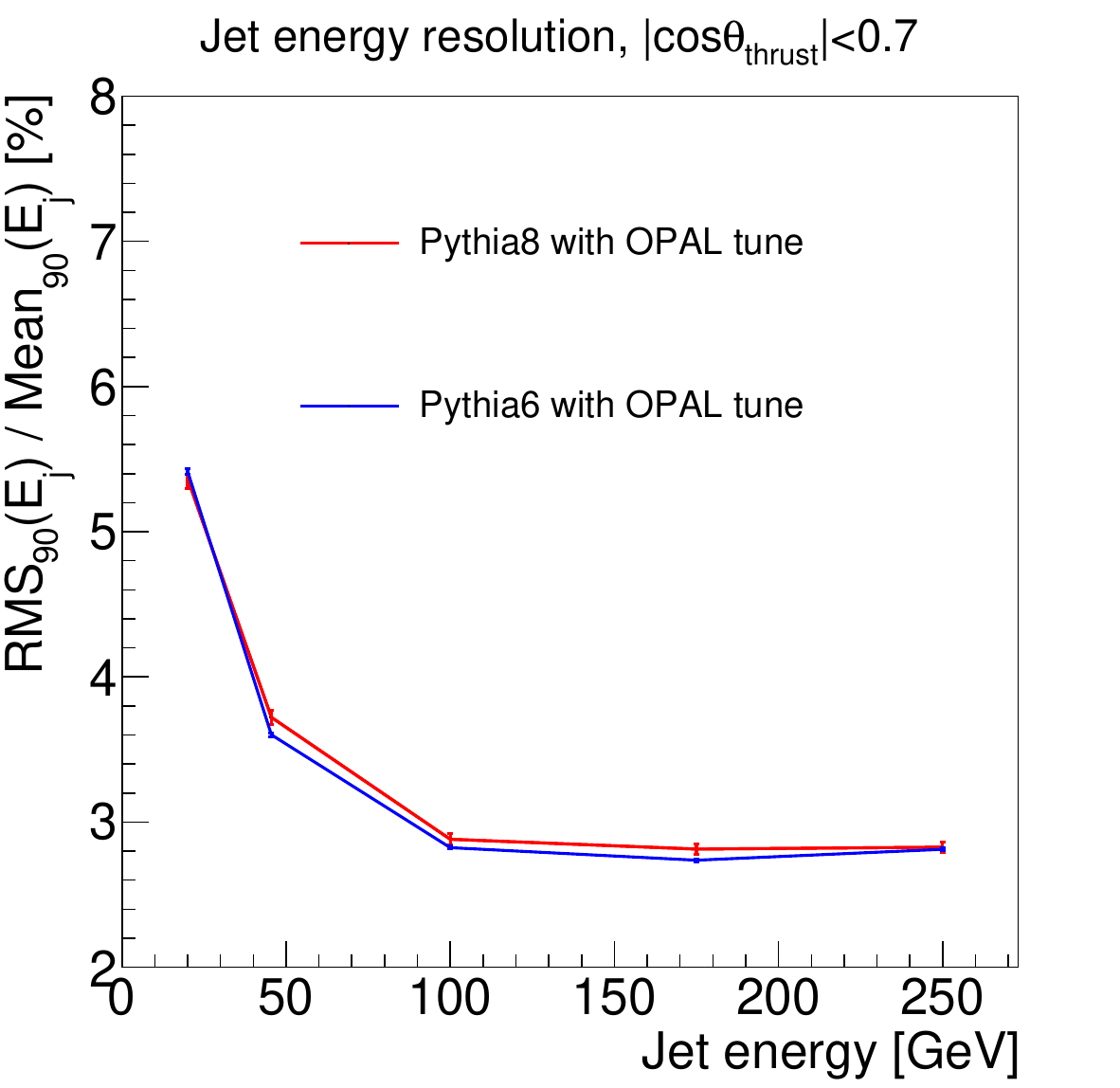}
  \end{subfigure}
  \begin{subfigure}[b]{0.32\linewidth}
    \includegraphics[width=\linewidth]{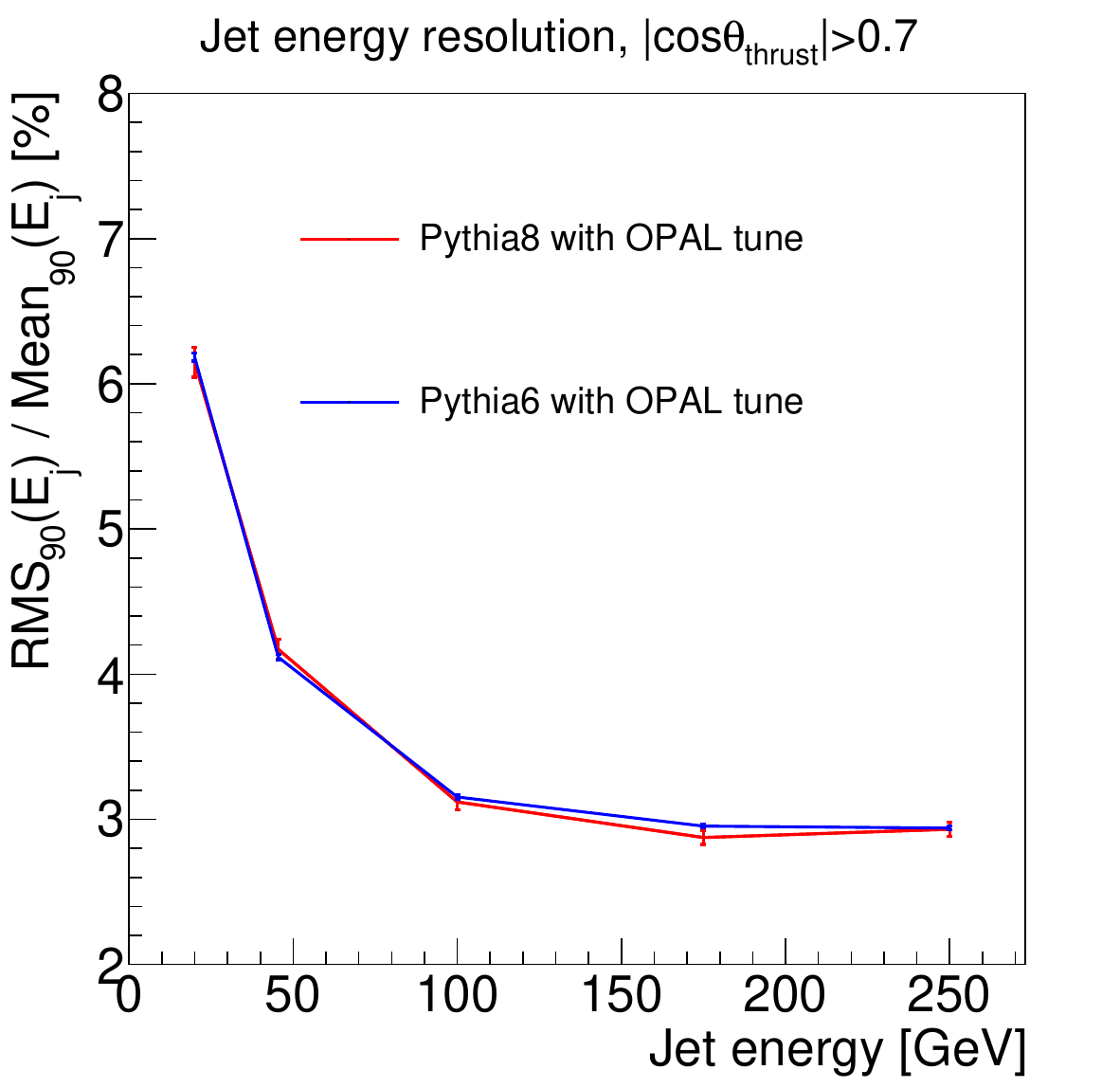}
  \end{subfigure}
  \caption{The JER of ILD are plotted at: left: full angle region, middle: barrel region, and right: forward region. Here, events are generated by two different Pythia versions with the OPAL tune.} \label{fig:jer1}
\end{figure}

As a first step, we reproduce the results in the ILD Interim Design Report (IDR)~\cite{ILDConceptGroup:2020sfq} by the Whizard3+Pythia8 framework.
Since the OPAL tune has been used in the IDR, 
we also generate events with the OPAL tune.
Comparison between our results and previous events is shown in Fig.~\ref{fig:jer1}. 
This figure is separated to 3 regions: 
left is the full angle region, 
middle is the barrel region ($|\cos\theta_{thrust}|<0.7$), 
and right is the forward region ($|\cos\theta_{thrust}|>0.7$). 
There are slight differences between two curves, 
but basically they have good agreement.

\begin{figure}[htbp]
  \centering
  \begin{subfigure}[b]{0.32\linewidth}
    \includegraphics[width=\linewidth]{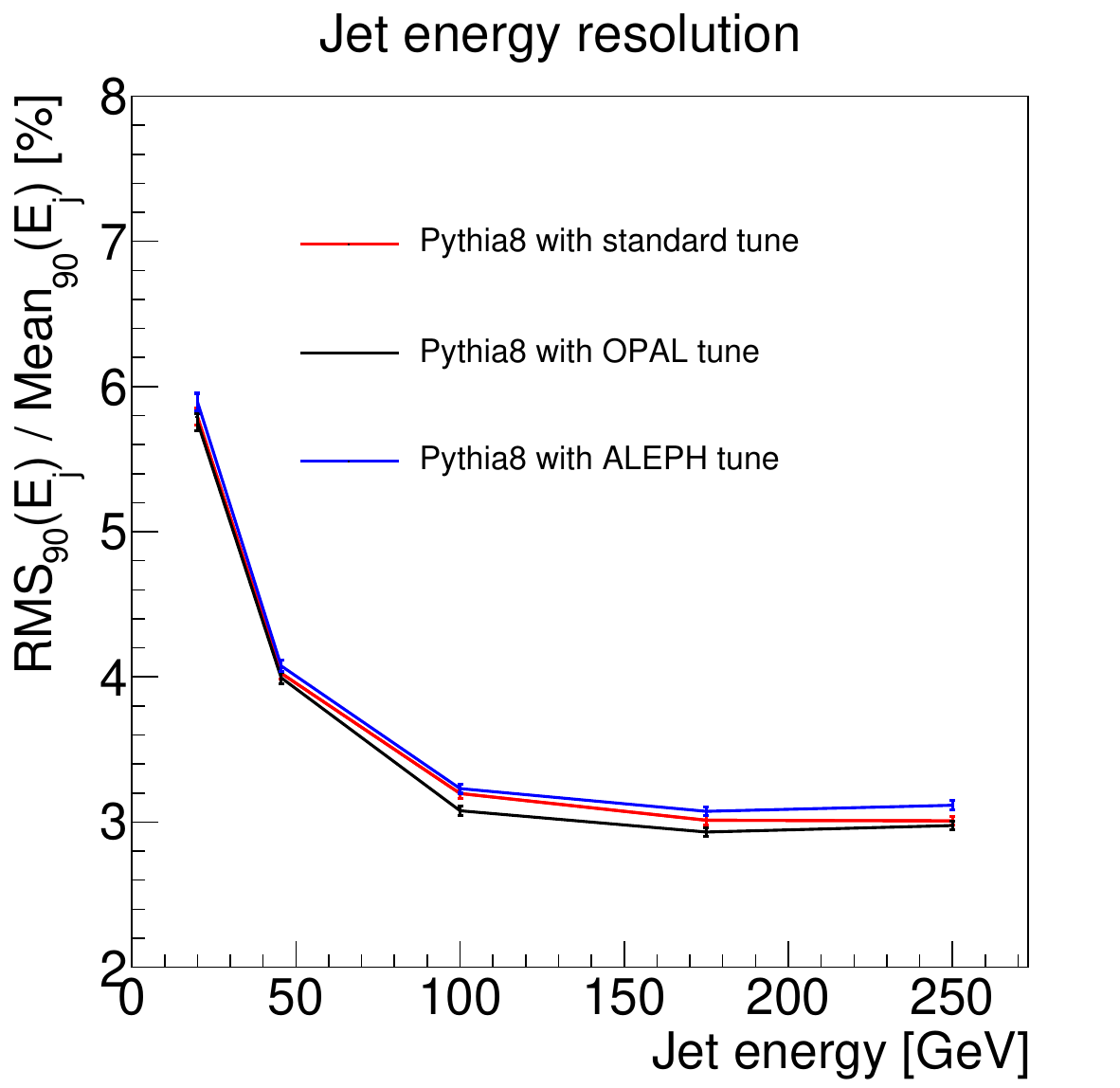}
  \end{subfigure}
  \begin{subfigure}[b]{0.32\linewidth}
    \includegraphics[width=\linewidth]{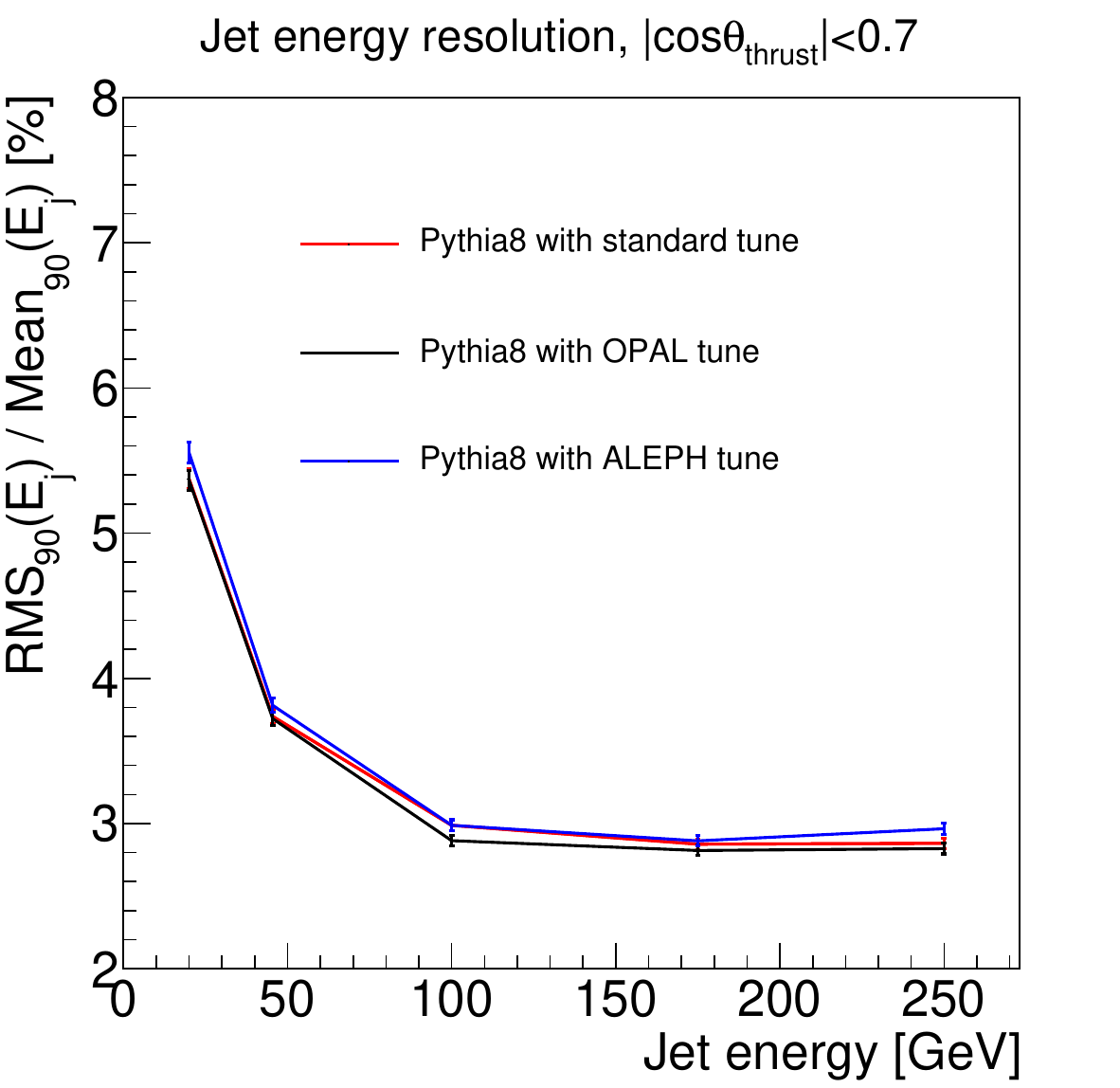}
  \end{subfigure}
  \begin{subfigure}[b]{0.32\linewidth}
    \includegraphics[width=\linewidth]{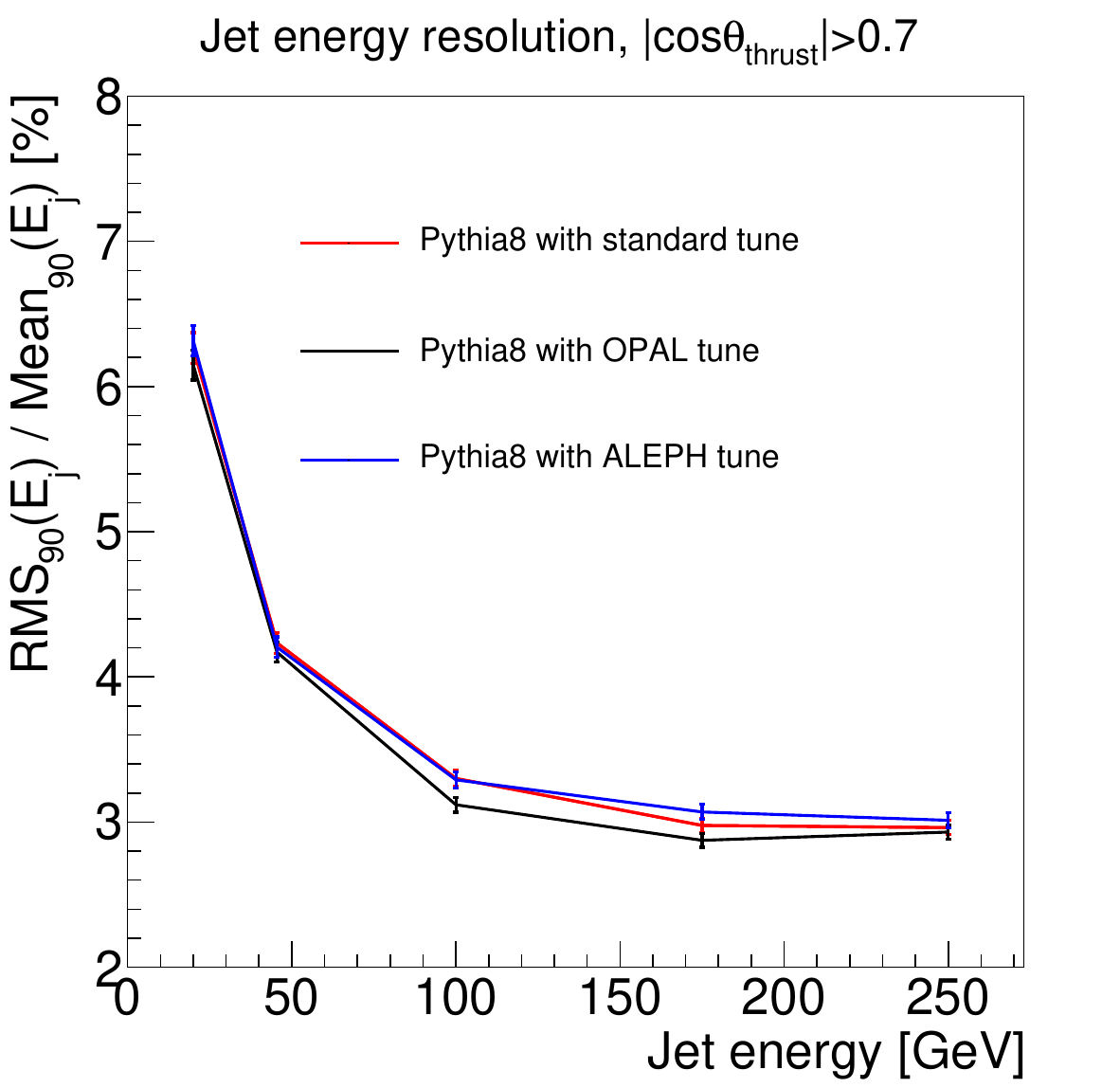}
  \end{subfigure}
  \caption{The JER of ILD are plotted at: left: full angle region, middle: barrel region, and right: forward region. Three Pythia8 tunes are compared: the standard, OPAL and ALEPH tunes. } \label{fig:jer2}
\end{figure}

We also compare the JER with differnt Pythia8 tunes. 
Results are shown in Fig.~\ref{fig:jer2}. 
All of them can give JER around 3\%. 
The standard tune and the ALEPH tune have some overlap, 
and the curve of OPAL tune is lower than other two curves.

\section{Preliminary Results of NLO Events}
\label{sec:nlo}

\begin{figure}[htbp]
  \centering
  \begin{subfigure}[b]{0.45\linewidth}
     \includegraphics[width=\linewidth]{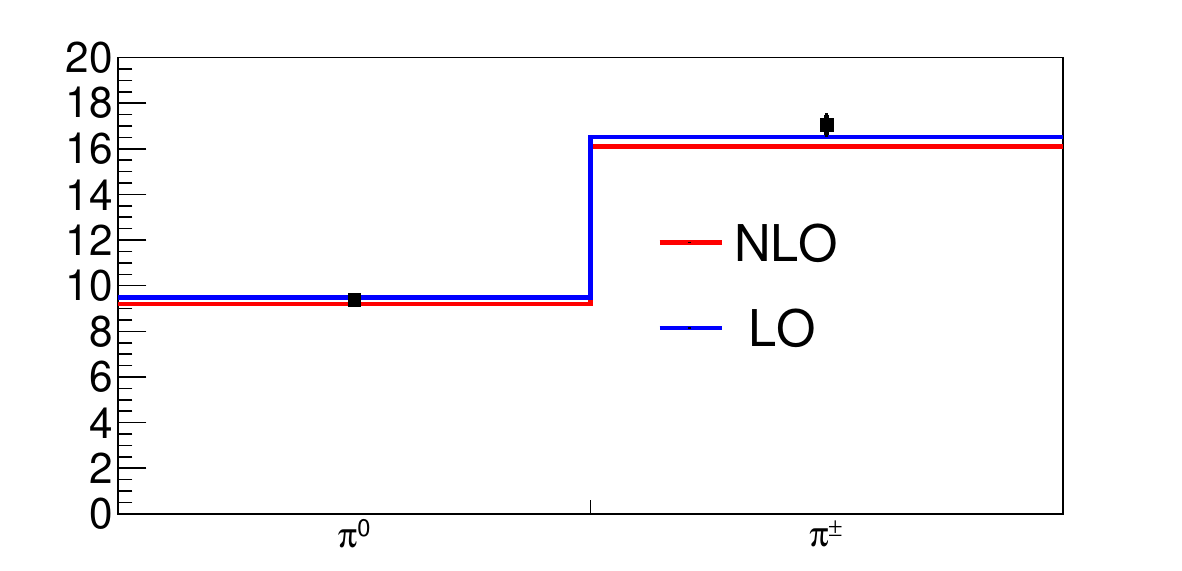}
  \end{subfigure}
  \begin{subfigure}[b]{0.45\linewidth}
     \includegraphics[width=\linewidth]{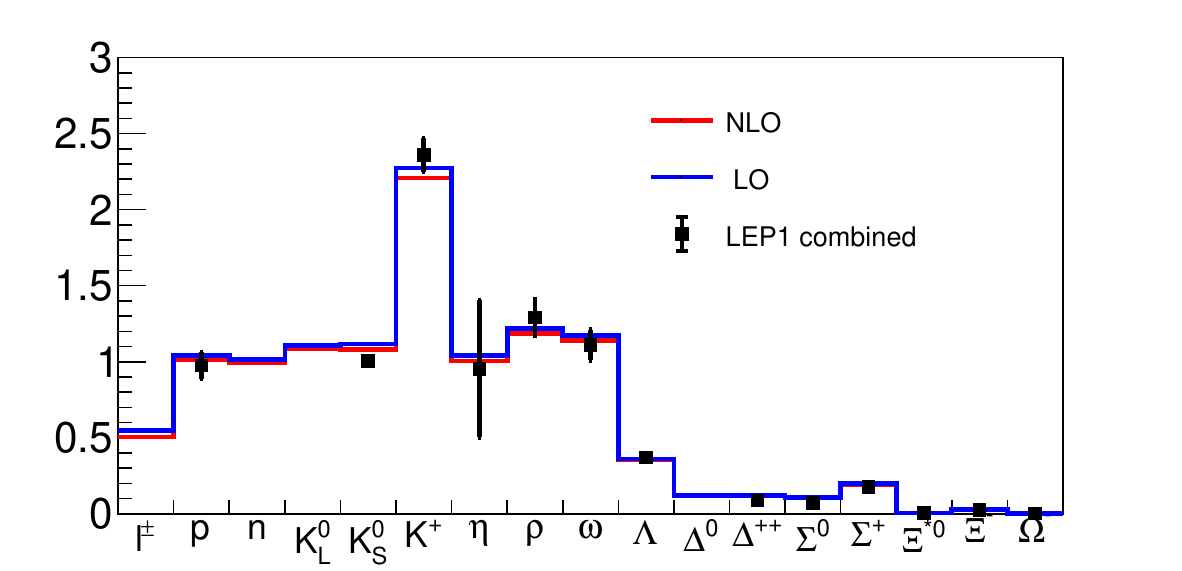}
  \end{subfigure}
  \caption{The average numbers of hadrons in $e^+e^-\to q\bar{q}$ events at LO and NLO are displayed, with the Pythia8 standard tune.} \label{fig:pions-nlo}
\end{figure}

NLO QCD corrections can be calculated by interfacing Whizard with OpenLoops~\cite{Buccioni:2019sur}.
Whizard supports POWHEG matching~\cite{Nason:2004rx} to generate NLO events. 
The average hadron multiplicities with Pythia8 standard tune are shown in Fig.~\ref{fig:pions-nlo}.
One can observe that the numbers of hadrons at NLO are slightly lower than the LO.

\begin{figure}[htbp]
  \centering
  \begin{subfigure}[b]{0.45\linewidth}
    \includegraphics[width=\linewidth]{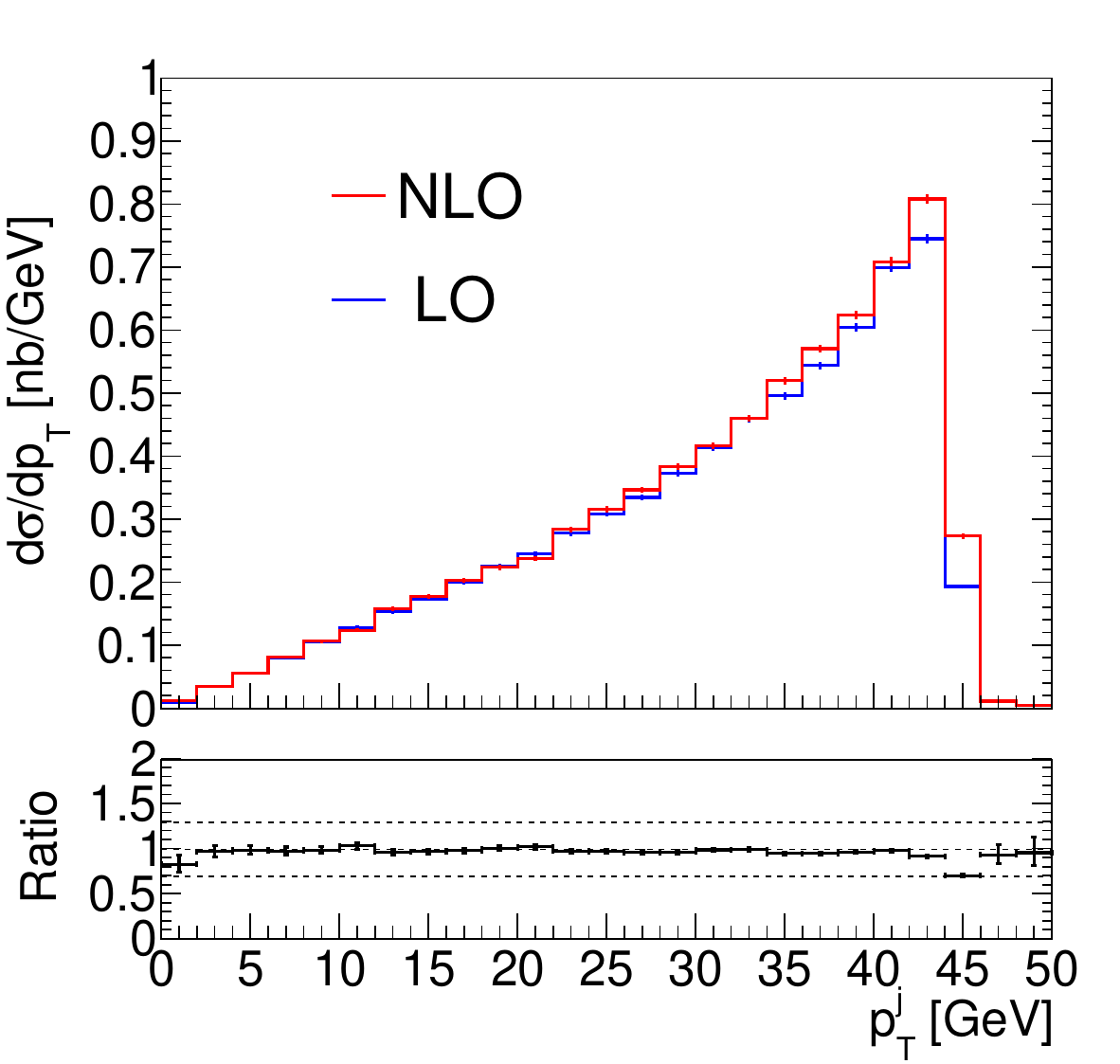}
    \subcaption*{Leading jet of $e^+e^- \to u\bar{u}$}
  \end{subfigure}
  \begin{subfigure}[b]{0.45\linewidth}
    \includegraphics[width=\linewidth]{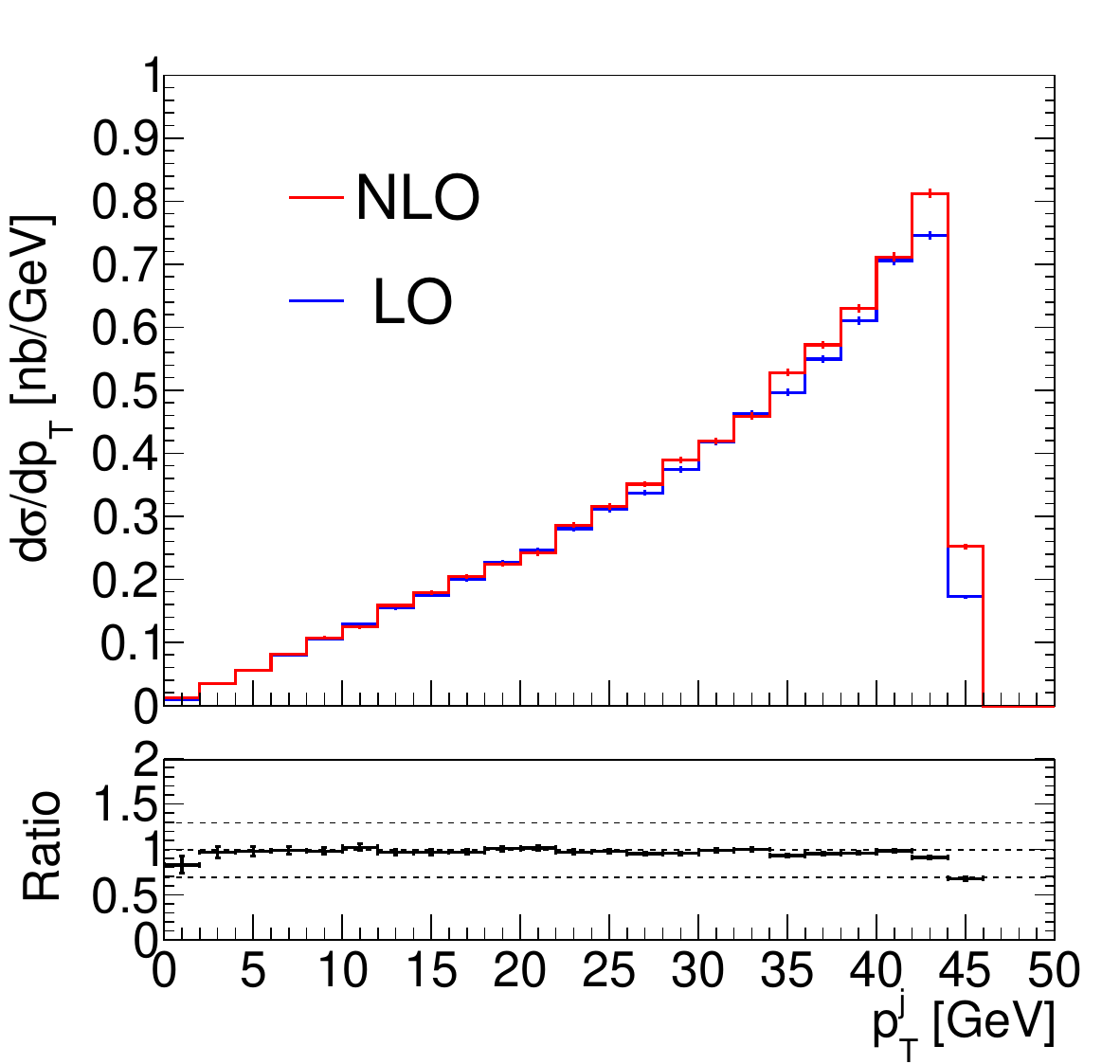}
    \subcaption*{Sub-leading jet of $e^+e^- \to u\bar{u}$}
  \end{subfigure}
  \caption{The transverse momentum distributions in $e^+e^-\to u\bar{u}$ events at LO and NLO are displayed, with the Pythia8 standard tune.} \label{fig:pt-nlo}
\end{figure}

To see the NLO effects, we use FastJet~\cite{Cacciari:2011ma} to find jets with the Durham algorithm~\cite{Catani:1991hj}.
The total number of jets is forced to 2. 
The transverse momentum distributions are displayed in Fig.~\ref{fig:pt-nlo}.
For the $e^+e^-\to u\bar{u}$ process, the QCD correction is small. 
We can see that the ratio is almost $1$ in these plots. 
Another feature is that the enhancement tends to occur at the large $p_T$ region.  

\begin{figure}[htbp]
  \centering
  \begin{subfigure}[b]{0.32\linewidth}
    \includegraphics[width=\linewidth]{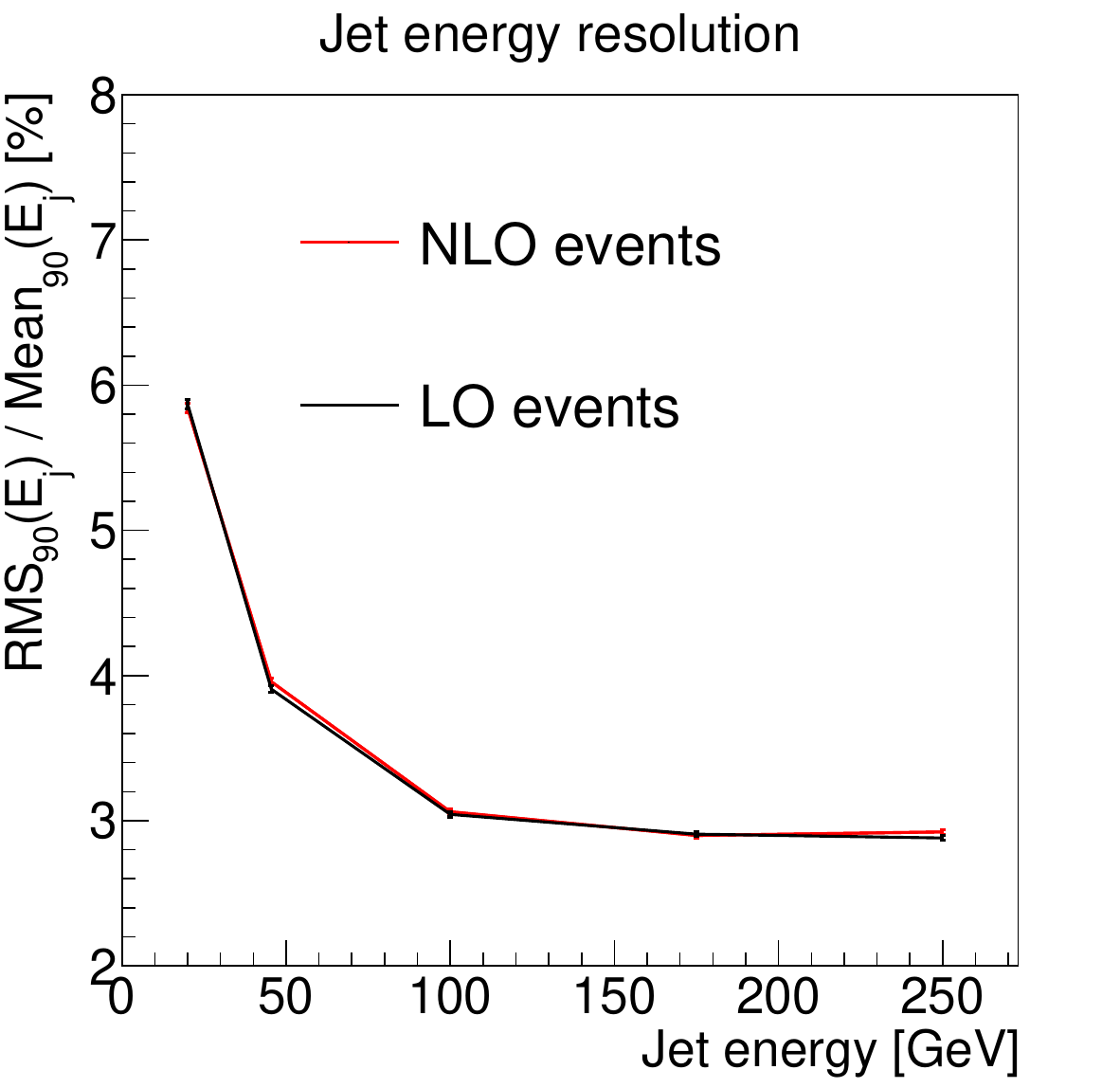}
  \end{subfigure}
  \begin{subfigure}[b]{0.32\linewidth}
    \includegraphics[width=\linewidth]{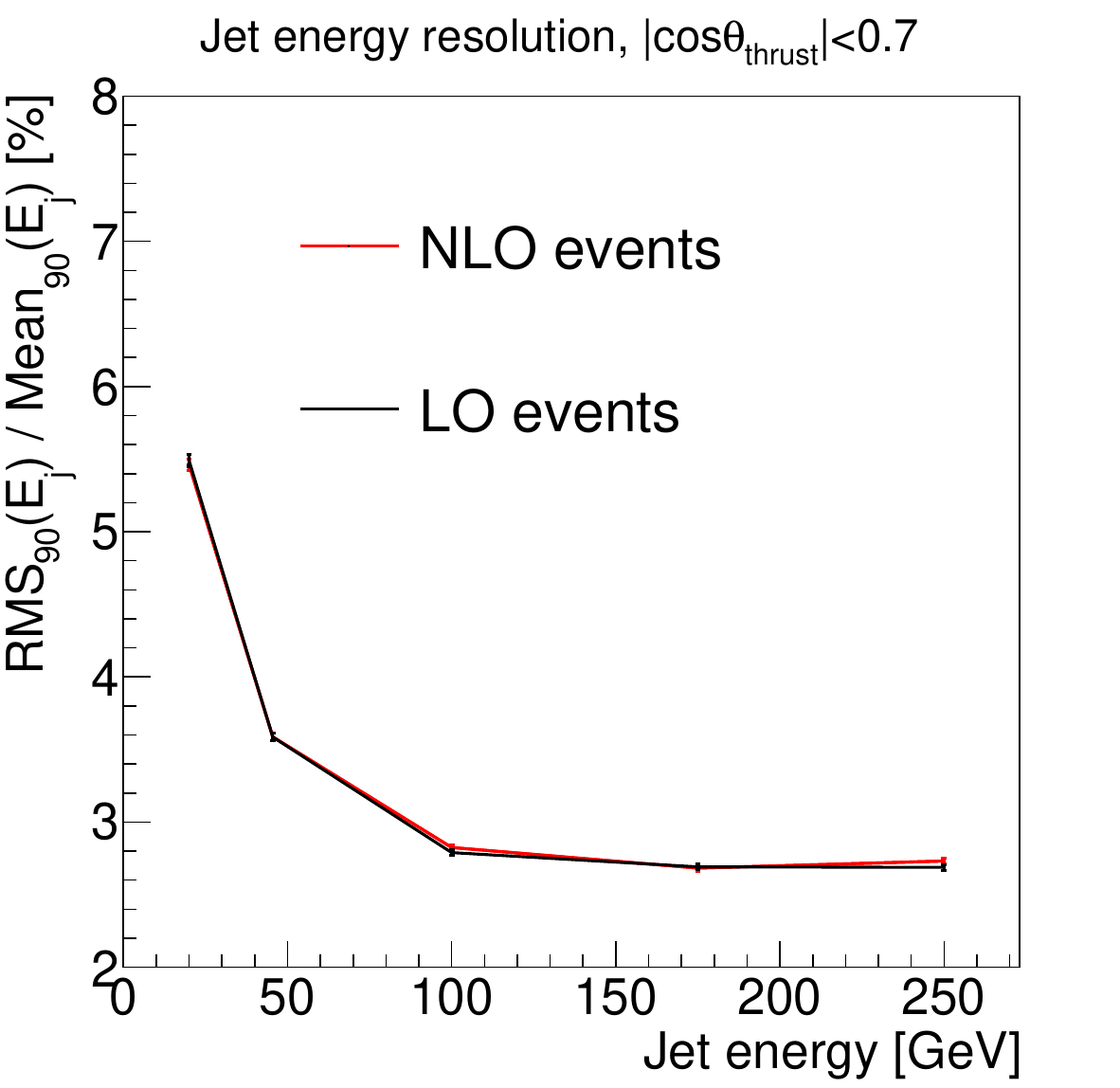}
  \end{subfigure}
  \begin{subfigure}[b]{0.32\linewidth}
    \includegraphics[width=\linewidth]{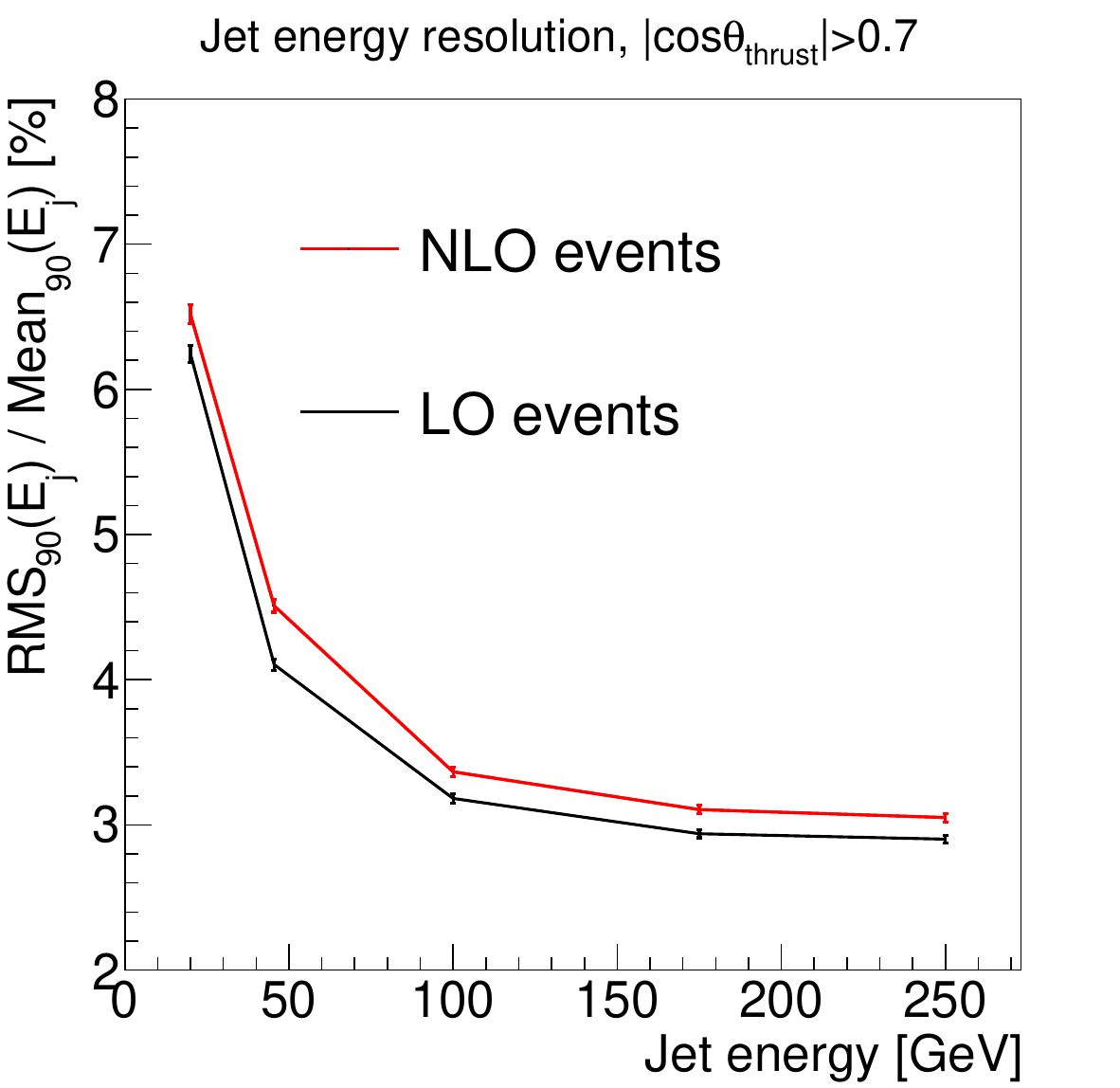}
  \end{subfigure}
  \caption{The JER of ILD are plotted at: left: full angle region, middle: barrel region, and right: forward region. Both the LO and NLO events are generated with the Pythia8 standard tune.} \label{fig:jer3}
\end{figure}

We can also pass the NLO events to perform an ILD simulation. 
The results of JER are plotted in Fig.~\ref{fig:jer3}. 
We can see that the JER curves at LO and NLO are consistent in the full angle region and the barrel region. 
But it is very different in the forward region. 

\begin{figure}[htbp]
  \centering
  \begin{subfigure}[b]{0.32\linewidth}
    \includegraphics[width=\linewidth]{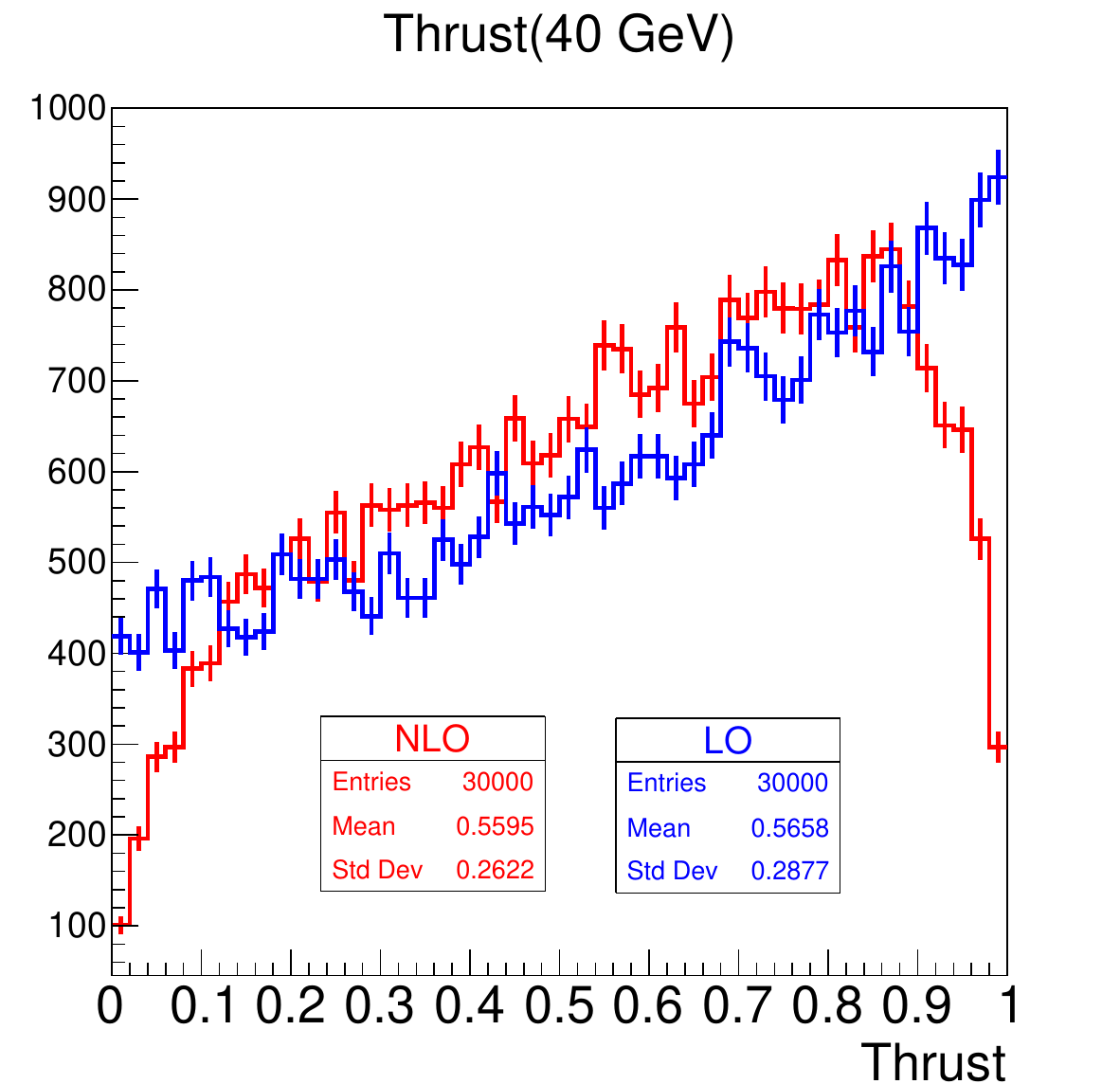}
  \end{subfigure}
  \begin{subfigure}[b]{0.32\linewidth}
    \includegraphics[width=\linewidth]{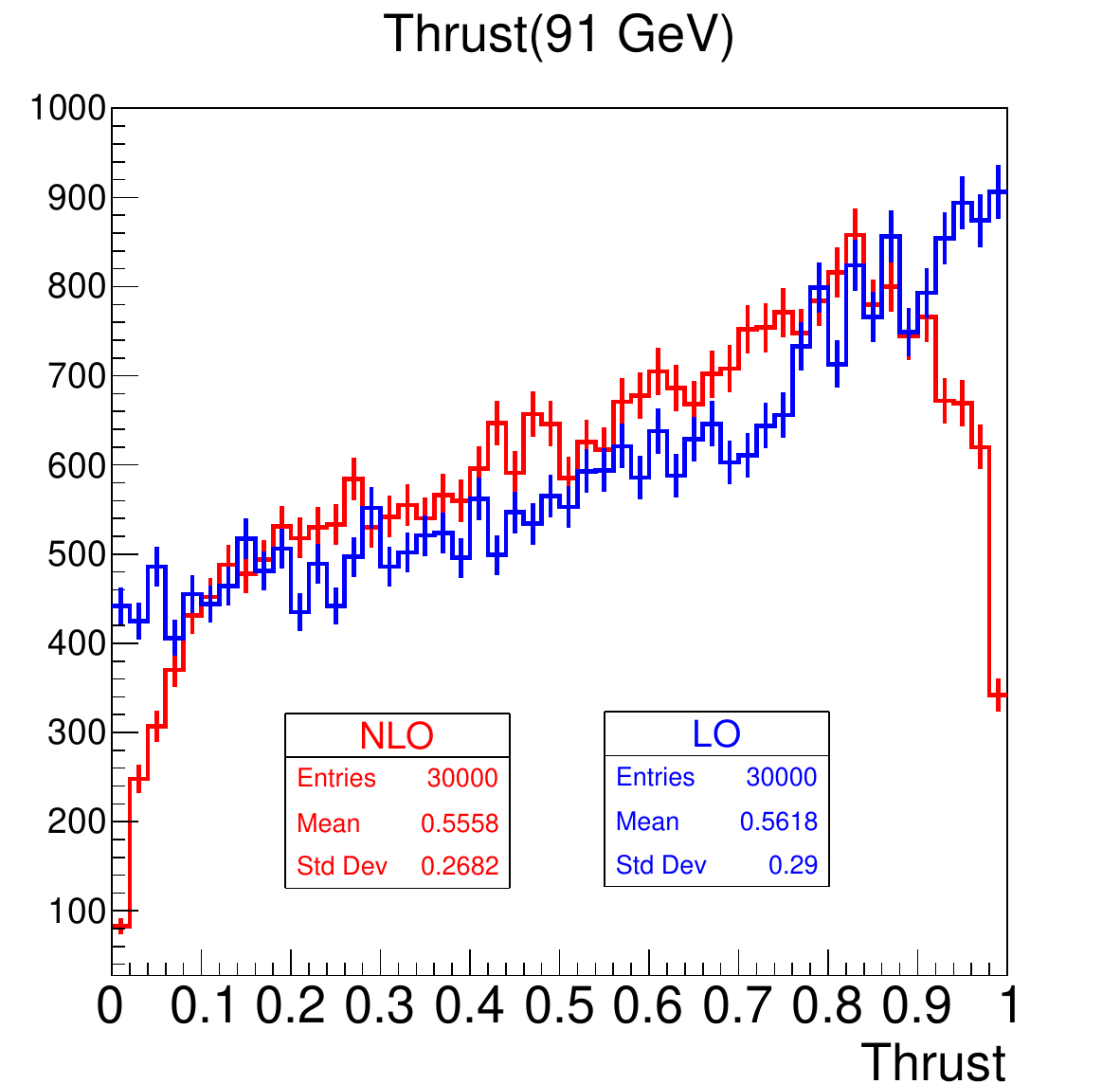}
  \end{subfigure}
  \begin{subfigure}[b]{0.32\linewidth}
    \includegraphics[width=\linewidth]{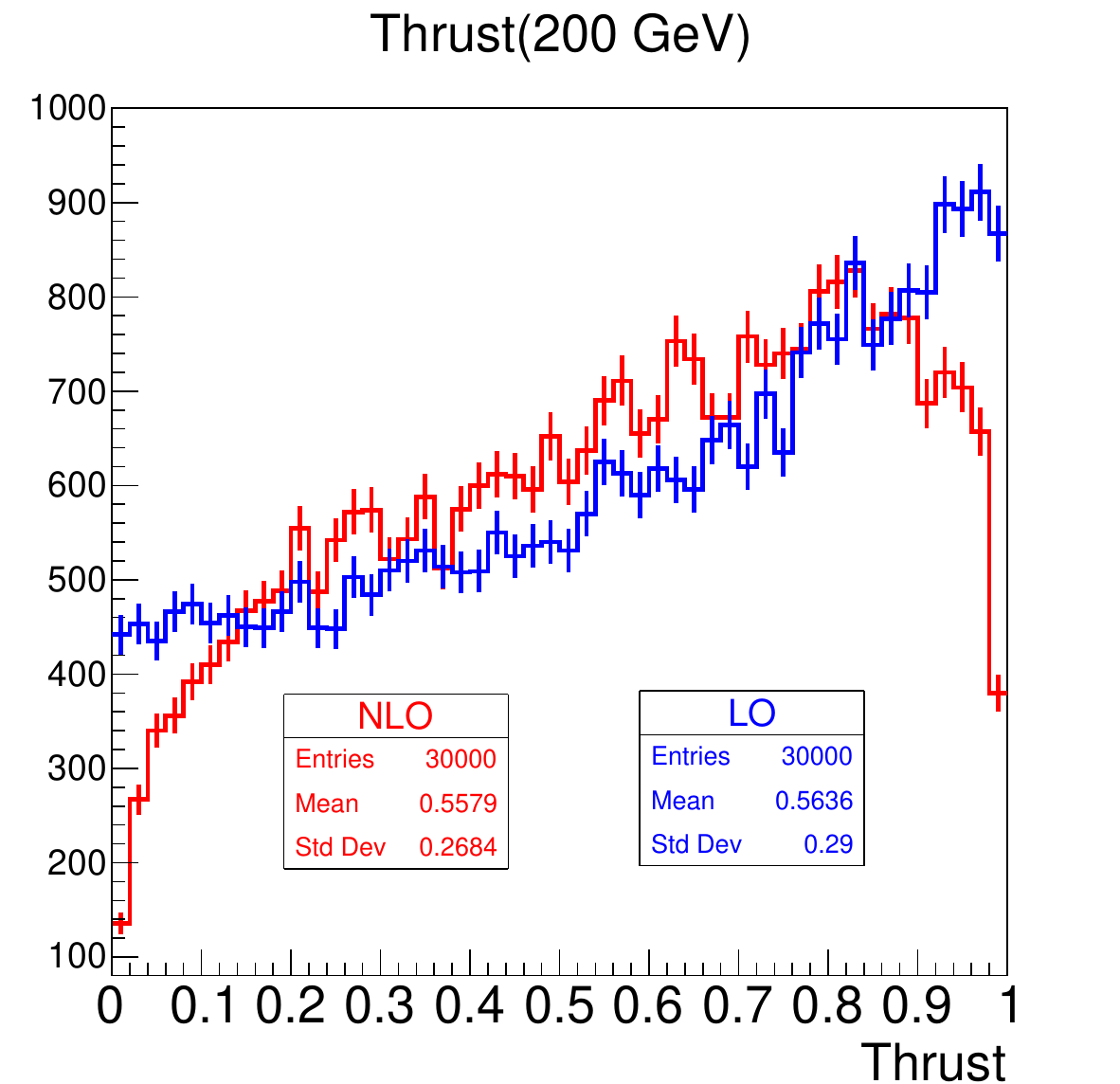}
  \end{subfigure}
  \begin{subfigure}[b]{0.32\linewidth}
    \includegraphics[width=\linewidth]{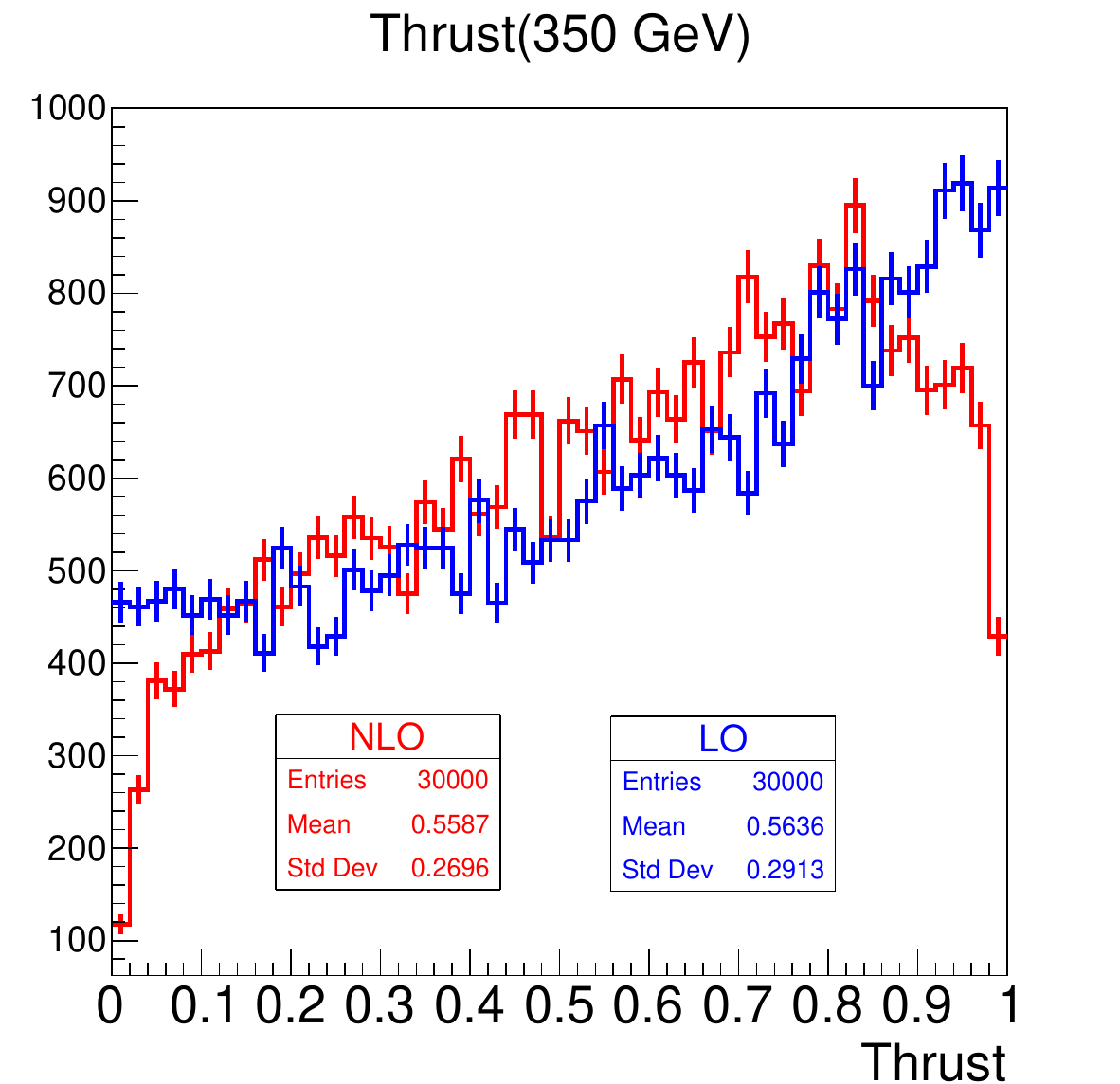}
  \end{subfigure}
  \begin{subfigure}[b]{0.32\linewidth}
    \includegraphics[width=\linewidth]{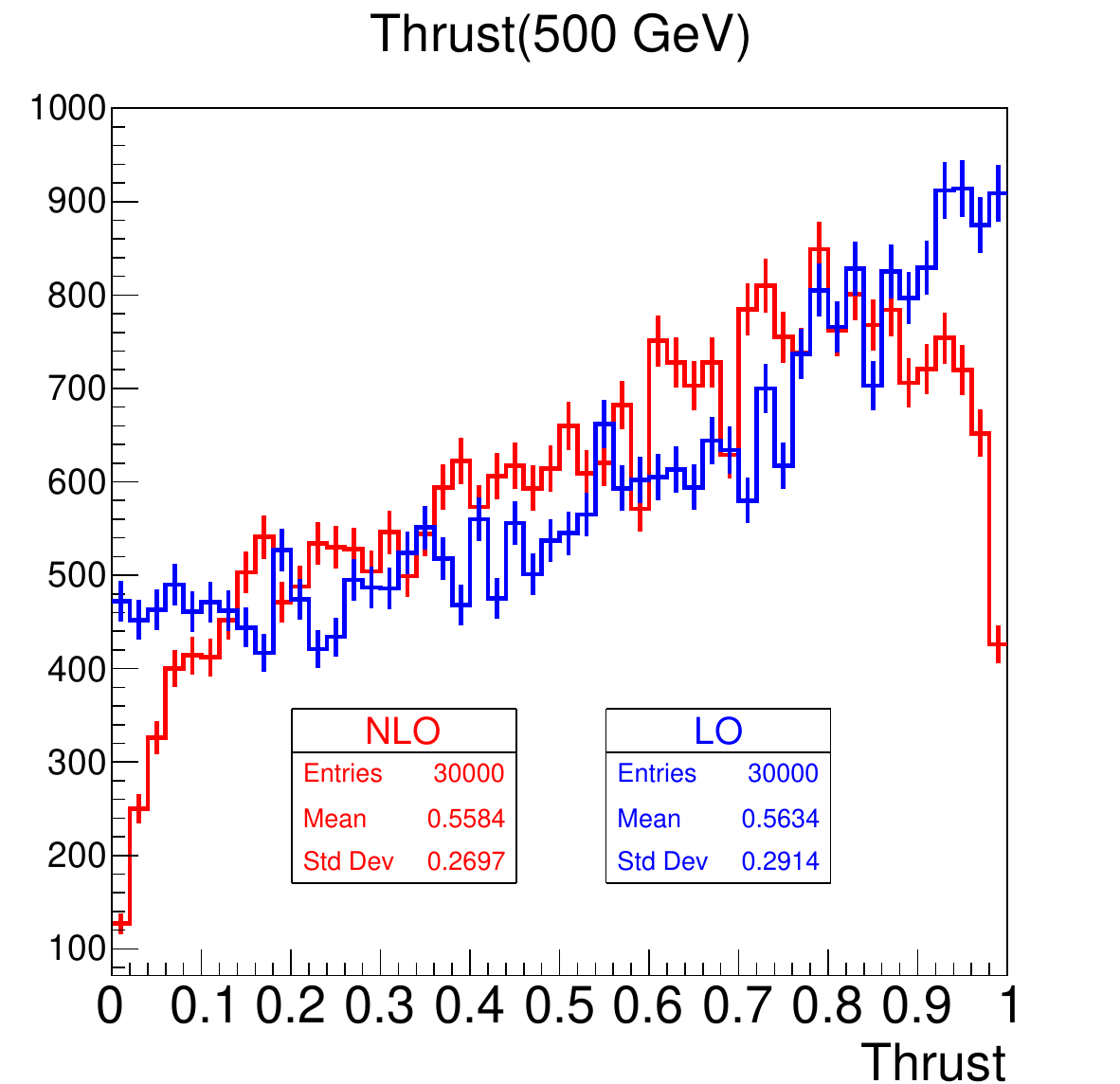}
  \end{subfigure}
  \caption{The thrust distributions at $E_{cm}=40,91,200,350,500$ GeV are displayed.} \label{fig:thrust}
\end{figure}

To figure out the reason, we plot the thrust distributions of two quarks at parton level in Fig.~\ref{fig:thrust}.
Obviously, there are fewer NLO events when the thrust tends to 0 or 1. 
In Whizard, there is a minimum $p_T$ for the hardest emission in an event. 
In our case, it is the additional gluon. 
This $p_T$ cut reduces the number of events in the forward region significantly.

\section{Summary}
\label{sec:summary}

In this contribution, we have upgraded the MC simulation chain for $e^+e^-$ colliders with the Whizard3+Pythia8 framework. 
As a first step, we study the process $e^+e^- \to q\bar{q}$.
Three tunes of Pythia8 have been compared: the standard Pythia8 tune, the OPAL tune and the ALEPH tune. 
The hadron multiplicities and JER at ILD have been compared. 
The standard tune can give hadron multiplicities close to LEP1 data, and the best JER is obtained by the OPAL tune. 
In this context, we are in favor of the Pythia8 standard tune, 
since it has better over-all agreement with the LEP1 data. 
We also test the NLO mode of Whizard and generate events by POWHEG matching. 
Preliminary results of NLO events are shown. 

\section*{Acknowledgements}

Z. Zhao has been partially supported by a China and Germany Postdoctoral Exchange Program between the Office of China Postdoctoral Council (OCPC) and DESY.

\section*{Appendix}

In this study, three tunes of Pythia8 are considered: 1) the standard tune with the default parameter set of Pythia8, 
2) the tune from the OPAL experiment, and 3) the tune from the ALEPH experiment.
The input parameters of these tunes are listed in Table.~\ref{beta1}.
The details of these parameters are referred to~\cite{Skands:2014pea}.

\begin{center}
\begin{table}
  \begin{center}
  \begin{tabular}{l|l|c|c|c}
  \hline
  Parameter      &  name in PYTHIA8  & standard   & OPAL   &  ALEPH\\
  \hline
  \hline
 $P(qq)/P(q)$  & StringFlav:probQQtoQ  & 0.081 &  0.085 &  0.105 \\
 $P(s)/P(u)$     & StringFlav:probStoUD  & 0.217  &  0.310 &  0.283 \\
 $(P(su)/P(du))/(P(s)/P(u))$        & StringFlav:probSQtoQQ & 0.915 & 0.45  &  0.710 \\
 $\frac{1}{3}(P(ud_1)/P(ud_0))$  & StringFlav:probQQ1toQQ0 & 0.0275  &  0.025  & 0.05 \\
 $(S=1)$ d,u   & StringFlav:mesonUDvector &  0.50  &  0.60 & 0.54 \\
 $(S=1)$ s      & StringFlav:mesonSvector    &  0.55   &  0.40 & 0.46 \\
 $(S=1)$ c,b   & StringFlav:mesonCvector   &  0.88   &  0.72  & 0.65 \\
                        & StringFlav:mesonBvector   &   2.20   &  0.72  &  0.65 \\
 $S=1,s=0$ prob. & StringFlav:mesonUDL1S0J1  &  0.0 &  0.43 & 0.12\\
 $S=0,s=1$ prob.  & StringFlav:mesonUDL1S1J0 & 0.0   &  0.08 & 0.04\\
 $S=1,s=1$ prob.  & StringFlav:mesonUDL1S1J1   &  0.0 &  0.08  & 0.12 \\
 tensor mesons (L=1)  & StringFlav:mesonUDL1S1J2 & 0.0 & 0.17 & 0.20 \\
 leading baryon suppr.  & StringFlav:suppressLeadingB & off  &  on  &  on\\
                                       & StringFlav:lightLeadingBSup   & 0.5 &  1.0 & 0.58 \\
                                      & StringFlav:heavyLeadingBSup & 0.9 &  1.0 & 0.58 \\
 $\sigma$ (GeV)      &  StringPT:sigma  &  0.335  &  0.4000 &  0.362 \\
 $\eta^\prime$ suppression  & StringFlav:etaPrimeSup  & 0.12  &  0.40  &  0.27\\ 
 $a$ of LSFF    &  StringZ:aLund & 0.68 &  0.11  &  0.40\\
 $b$ of LSFF     &  StringZ:aLund & 0.98  &  0.52 & 0.824 \\
 $\Delta a$ for s quark & StringZ:aExtraSQuark &  0.0 &0.0 & 0.0   \\
 $\Delta a$ for Di-quark & StringZ:aExtraDiquark & 0.97 & 0.5 & 0.5 \\
 $\epsilon_c$    &  StringZ:usePetersonC  & off  &  on  &   on\\
                           &  StringZ:epsilonC  & 0.05        &  $-0.031$ &  0.04\\
 $\epsilon_b$    &  StringZ:usePetersonB   & off  &  on  &  on \\
                           &  StringZ:epsilonB  & 0.005       &  $-0.002$ &  0.0018\\
 PS QCD cut-off (GeV)  & TimeShower:pTmin  &  0.5    & 0.95  &  0.735  \\
 PS cut-off for QED  & TimeShower:pTminChgQ  &  0.5    & 0.95  &  0.735  \\
 adiation off quarks (GeV) &  &  & &  \\
  \hline
  \hline
  \end{tabular}
  \end{center}
  \caption{The input parameters of three tunes of Pythia8: the standard, OPAL and ALEPH tunes are listed here.} \label{beta1}
\end{table}
\end{center}

\printbibliography{}
\end{document}